\documentclass[aip,apl,amsmath,amssymb,reprint]{revtex4-2}

\usepackage[makeroom]{cancel}
\usepackage{dcolumn}
\usepackage{makecell}
\usepackage[utf8]{inputenc}
\usepackage[T1]{fontenc}
\usepackage{mathptmx}
\usepackage{commath}
\usepackage{graphicx}
\usepackage{multirow}
\usepackage{amsmath}

\usepackage{soul}
\usepackage{verbatim}
\usepackage{bm}
\usepackage{outlines}
\usepackage{siunitx}
\usepackage[space]{grffile}
\usepackage[dvipsnames]{xcolor}
\usepackage{xr}
\usepackage{color}
\usepackage{tikz}
\usepackage{mathtools}
\usepackage{siunitx}
\usepackage{amssymb,placeins}
\usepackage[globalcitecopy]{bibunits}
\usepackage{layouts}

\newcommand\myshade{85}
\colorlet{mylinkcolor}{violet}
\colorlet{mycitecolor}{YellowOrange}
\colorlet{myurlcolor}{Aquamarine}
\usepackage{hyperref}
\hypersetup{
  linkcolor  = mylinkcolor!\myshade!black,
  citecolor  = mycitecolor!\myshade!black,
  urlcolor   = myurlcolor!\myshade!black,
  colorlinks = true,
}
\usepackage[capitalize]{cleveref}
\usetikzlibrary{calc, positioning}

\newcommand{\kOhm}{\mathrm{k}\Omega}
\newcommand{\Celsius}{^\circ\mathrm{C}}
\newcommand{\mm}{\mathrm{mm}}
\newcommand{\um}{\mu \mathrm{m}}
\newcommand{\nm}{\mathrm{nm}}
\newcommand{\seconds}{\mathrm{s}}
\newcommand{\minutes}{\mathrm{min}}
\newcommand{\GHz}{\mathrm{GHz}}
\newcommand{\MHz}{\mathrm{MHz}}
\newcommand{\dBm}{\mathrm{dBm}}
\newcommand{\Watt}{\mathrm{W}}
\newcommand{\mTorr}{\mathrm{mTorr}}
\newcommand{\QuTech}{\affiliation{QuTech and Kavli Institute of Nanoscience, Delft University of Technology, P.O. Box 5046, 2600 GA Delft, The Netherlands}}

\begin{document}

\title{Lower-temperature fabrication of airbridges by grayscale lithography to increase yield of nanowire transmons in circuit QED quantum processors}
\author{T.~Stavenga}\QuTech
\author{L.~DiCarlo}\QuTech

\date{\today}

\begin{abstract}
Quantum hardware based on circuit quantum electrodynamics makes extensive use of airbridges to suppress unwanted modes of wave propagation in coplanar-waveguide transmission lines. Airbridges also provide an interconnect enabling transmission lines to cross. Traditional airbridge fabrication produces a curved profile by reflowing resist at elevated temperature prior to metallization. The elevated temperature can affect the coupling energy and even yield of pre-fabricated Josephson elements of superconducting qubits, tuneable couplers and resonators. We employ grayscale lithography in place of reflow to reduce the peak airbridge processing temperature from $200$ to $150\Celsius$, showing a substantial yield increase of transmon qubits with Josephson elements realized using Al-contacted InAs nanowires.
\end{abstract}
\maketitle

\begin{bibunit}[apsrev4-2]

Free-standing metallic strips bridging separate planar conductors, called airbridges (ABs)~\cite{Koster89}, are widely used in classical~\cite{Lankwarden12} and quantum~\cite{Abuwasib13, Chen14, Riste15, Janzen22} microwave-frequency integrated circuits. They are most commonly employed to suppress slotline-mode wave propagation in coplanar-waveguide transmission lines (CPWs)~\cite{Wen69, Simons01} by connecting the ground planes flanking the central conductor, thereby avoiding spurious resonance modes and reducing crosstalk. A second use of ABs is as interconnect allowing transmission lines to cross with low impedance mismatch and crosstalk.

ABs are intensely used in superconducting quantum hardware based on circuit QED~\cite{Blais04, Krantz19}, where CPWs are commonly used to make resonators for qubit readout and qubit-qubit coupling, as well as qubit control lines. For example, in our planar quantum hardware architecture~\cite{Versluis17} designed for surface-code error correction, 7- and 17-qubit processors contain $\sim 600$  and $\sim 1200$ ABs, respectively, of which 3 and 20  are used for crossovers~\cite{Marques22}. In the 49-qubit version, the number of AB crossovers jumps to 130 owing to the routing of qubit control lines from the chip periphery to more qubits at the center. Signal routing at higher qubit counts requires advanced methods based on three-dimensional integration, including  through-silicon vias~\cite{Alfaro_Barrantes20, Yost20, Mallek21}, bump bonding~\cite{Rosenberg17,Foxen17}, and the chip packaging itself~\cite{Bronn18}. In this context, ABs remain essential for slotline-mode suppression and crossovers.

ABs are typically added in the final fabrication step as otherwise resist non-uniformity induced by the few-$\um$ height of ABs can reduce yield and increase variability of post-fabricated circuit elements
(for exceptions, see Refs.~\cite{Andersen20, Krinner22}).  The most traditional AB fabrication method uses resist reflow at elevated temperature to produce ABs with smooth, rounded profile.
However, many types of Josephson junctions (JJs) are not compatible with this elevated temperature.
Examples include the semiconductor-normal-superconductor (SNS) JJs based on InAs~\cite{Krogstrup15} and InSb nanowires~\cite{Heedt21} used in SNS transmons~\cite{Larsen15, Luthi18} (also called gatemons and nanowire transmons).
The temperature excursions can reduce JJ yield at worst and unpredictably affect the JJ coupling energy at best, affecting qubit frequency targeting.

In this Letter, we apply grayscale lithography (GSL), a method most commonly used to fabricate microlenses~\cite{Loomis16, Deng17, Mortelmans20}, to reduce the peak AB processing temperature from $200\Celsius$ (required for standard reflow) to $150\Celsius$ (limited by resist adhesion). We detail our calibration of GSL to accurately produce a curved resist-height profile by spatial control of electron-beam (e-beam) resist dose, with pre-compensation for proximity effect and resist nonlinearity.  Our main result is the demonstration that the reduction in peak processing temperature increases the yield of SNS transmons with junctions realized using epitaxially grown, Al-contacted InAs nanowires. Very recent work~\cite{Janzen22} has demonstrated the use of GSL to fabricate ABs with a single e-beam step, showing compatibility with transmons based on standard superconductor-insulator-superconductor (SIS) JJs. Our focus here is on SNS JJ compatibility, with emphasis on the positive impact of AB fabrication at lower peak temperature as enabled by GSL.

\begin{figure}
\centering
\includegraphics{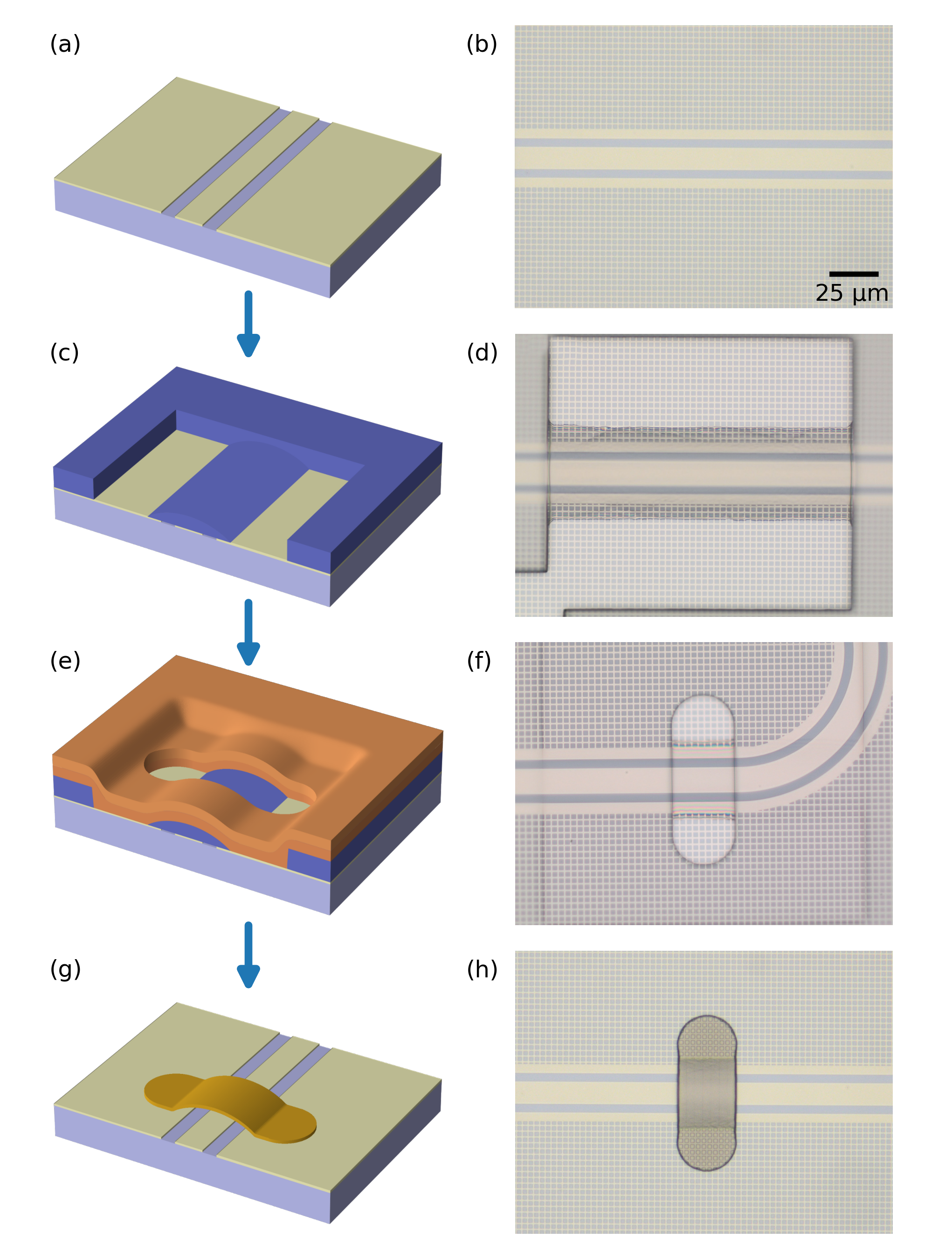}
\caption{Overview of airbridge fabrication by the GSL method, using (left) schematics and (right) optical images.
(a,b) Pre-fabrication of the base layer. Our CPW transmission lines have $12~\um$ center conductor width and $4~\um$ gaps between the central conductor and the flanking ground planes.  (c,d) Patterning of the PMGI (blue) bottom resist layer using GSL. (e,f) Patterning of the PMMA top resist bilayer (orange) defining the lateral dimensions of airbridges. (g, h) Sputtering of NbTiN (gold) and liftoff.
\label{fig:Fab_process}
}
\end{figure}

\begin{figure}
\centering
\includegraphics{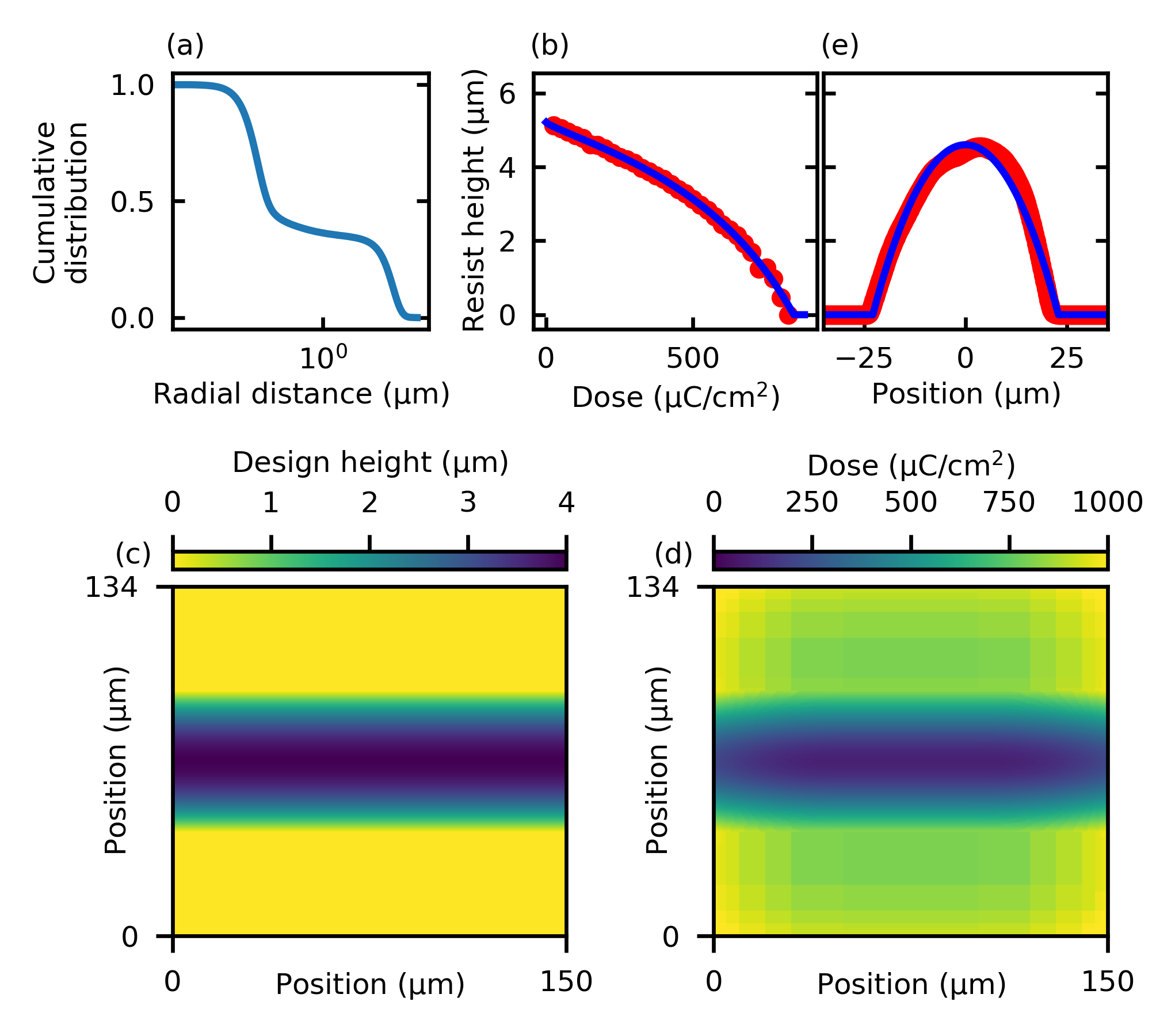}
\caption{
Calibration of grayscale e-beam lithography.
(a) CDF of the energy of the e-beam in PMGI on top of NbTiN. Note that more than $30\%$ of the energy is deposited beyond a $20~\um$ radius.
(b) Calibration of PMGI height as a function of local e-beam dose (red) and fit (blue) used for interpolation by the software.
(c) Two-dimensional image of the targeted resist height for the airbridge.
(d) Image of the dose map required to achieve the height map in (d) with precompensation for proximity effect and resist nonlinearity.
(e)  Vertical line cut (red) of actual PMGI resist height as measured with a profilometer and best fit to a circle function (blue).
\label{fig:Grayscale_lithography}
}
\end{figure}

AB fabrication by GSL (Fig.~\ref{fig:Fab_process}) starts after defining the chip base layer containing all CPW structures and transmons, including their SNS junctions. A layer of PMGI (blue) SF15 ($6.4$ or $3~\um$ thick, see below) is spun and baked for $5~\minutes$ on a hotplate at $150\Celsius$. This is found to be the lowest viable temperature avoiding resist adhesion problems. Using e-beam lithography and GSL, the AB profile and clearances are then written. An AZ400K/water mixture in a 1:4 volume ratio is used for development. The chip is dunked into the developer for $35~\seconds$ followed by a thorough water rinse for $30~\seconds$ and  blow-drying. At this point, we typically check for correctness by measuring the height profile along the curve of an AB using a profilometer [Fig.~\ref{fig:Grayscale_lithography}(c)]. Next, a $400~\nm$ thick layer of PMMA 495K (orange) is spun and baked in a vacuum oven at $100\Celsius$ for $10~\minutes$, immediately followed by a $1.5~\um$ thick layer of PMMA 950k (orange) spun and baked in the same way. E-beam lithography and resist development define the lateral dimensions of the ABs. The top-layer resists must be compatible with the bottom-layer resist. This means that the top layer solvent cannot dissolve the bottom resist after it has been developed and that the developer for the top layer resists cannot develop the bottom layer. A $30~\seconds$ buffered oxide etch with 1:1 dilution factor is performed prior to metal deposition. We next sputter $200~\nm$ of NbTiN (gold) without any argon milling as the plasma can induce currents in the SNS junctions, causing their failure. A photoresist, $700~\nm$ of S1805 baked at $85\Celsius$ for $3~\minutes$, is used for protection during dicing. After dicing, this resist is lift-off using $88\Celsius$ N-methyl pyrrolidone (NMP) for $15~\minutes$ and followed by two rinses in isopropanol (IPA) at $80\Celsius$  for $10~\minutes$. Due to the conformal nature of sputtering, there is a vertical edge of NbTiN left that is approximately the height of the bottom PMMA layer.

\begin{figure}
\centering
\includegraphics{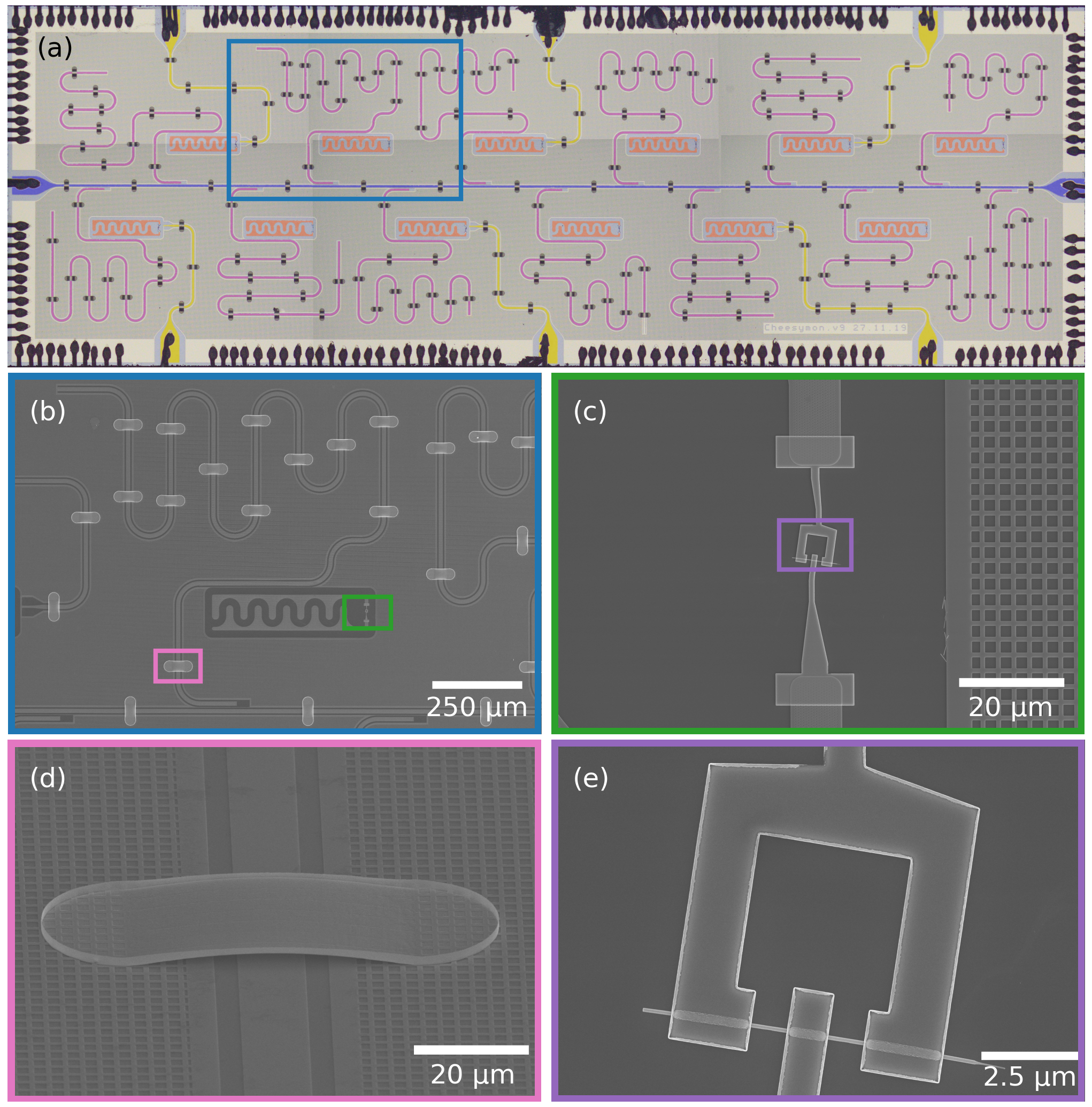}
\caption{
Images at various length scales of a circuit QED test device with $100\%$ yield of 185 airbridges fabricated by the GSL method.
(a) Optical image of the full device  $(7~\mm\ \times \ 2.3~\mm)$, with added falsecolor. The device has 12 flux-tuneable SNS transmons (red) with dedicated readout resonators (purple) coupled to a common readout feedline (blue). Six of the SNS transmons have dedicated flux-control lines (yellow).
(b,e) Scanning electron micrographs (SEM) showing (b) one SNS transmon and its dedicated readout resonator; (c) the SNS junction pair and its connection to the transmon capacitor pads; (e) zoom-in on the SNS junction pair and SQUID loop; and (d) an example airbridge.
\label{fig:Nanowire_transmon}
}
\end{figure}

Figure~\ref{fig:Nanowire_transmon} shows a complete circuit QED test device with 185 ABs fabricated by GSL and with $100\%$ yield.
The device consists of 12 flux-tuneable SNS transmons each with a dedicated readout resonator coupling to a common feedline. Six of the transmons have dedicated flux bias lines, but all can be globally tuned using an external coil. The flux-tuneable Josephson element in each transmon consists of two Al/InAs/Al junctions in parallel with loop area $\sim20~\um^2$. The two junctions are fabricated from a common hexagonal InAs nanowire with $100~\nm$ diameter and two facets covered with epitaxially grown Al ($10~\nm$ thick). Each SNS junction is defined by etching a $\sim200~\nm$ section of Al [Fig.~\ref{fig:Nanowire_transmon}(e)].

Contrary to the traditional method of producing a curved AB profile by reflowing the PMGI at elevated temperature $(200\Celsius)$, GSL achieves the rounding by spatial control of the e-beam dose. For a positive resist like PMGI, a lower (higher) dose causes slower (faster) removal of the resist, resulting in a higher (lower) remnant resist thickness. Our desired resist-height profile is semi-circular, mimicking the profile achieved in the reflow process by surface tension. To achieve this, it is necessary to correct for proximity error as long-range scattering deposits up to $30\%$ percent of the e-beam energy at a range exceeding $20~\um$ [Fig.~\ref{fig:Grayscale_lithography}(a)]. If this effect is not compensated, areas with dense (sparse) features are overexposed (underexposed). It is also important to  calibrate the non-linear dose-height correspondence (contrast curve). Non-lineariy is desirable in typical microfabrication, as almost all processes require a binary resist profile (so-called perfect contrast) in which the resist is either not exposed or fully exposed. On the other hand, a linear resist is ideal for GSL. The non-linearity of PMGI ($6.4~\um$ thick) is evident in the measured contrast curve shown Fig.~\ref{fig:Grayscale_lithography}(b). We precompensate proximity and resist nonlinearity using the three-dimensional proximity effect correction (3D-PEC) module in the GenISys BEAMER software~\cite{Beamer}. The inputs are the point spread function of the energy deposited by the e-beam lithography machine on the resist stack, the interpolated contrast curve~\cite{Mack87} and the desired height map [Fig.~\ref{fig:Grayscale_lithography}(c)]. The output is a prescribed position-dependent dose. Following these calibrations, we actually reduced the thickness of the PMGI layer to $3~\um$ in order to reduce stress in the film, which at the original thickness caused cracks in the resist and many nanowires to detach. By reducing the development time from $50$ to $30~\seconds$, the calibrations were found to remain valid. This GSL process has very high yield and is stable with time. The first and last fabrication runs performed using the process, 16 months apart, yielded very similar airbridges without recipe adjustments.

GSL avoids the PMGI reflow step needed in the traditional method, reducing the peak PMGI temperature from $200\Celsius$ to $150\Celsius$.
We devise a simplified test to investigate the effect of PMGI peak temperature on SNS JJ room-temperature resistance.
This test entails spinning $3~\um$ of PMGI on two chips with arrays of single junctions. Next, one chip is heated on a hotplate for $5~\minutes$ to $150\Celsius$ while the other is heated to $200\Celsius$. The chips are not directly placed on the hotplate; rather, as is standard practice, a Si wafer (6" diameter) is placed in between. Finally, the resist is stripped off using a bath of NMP at $88\Celsius$ followed by two baths of IPA at $80\Celsius$.

\begin{figure}
\centering
\includegraphics{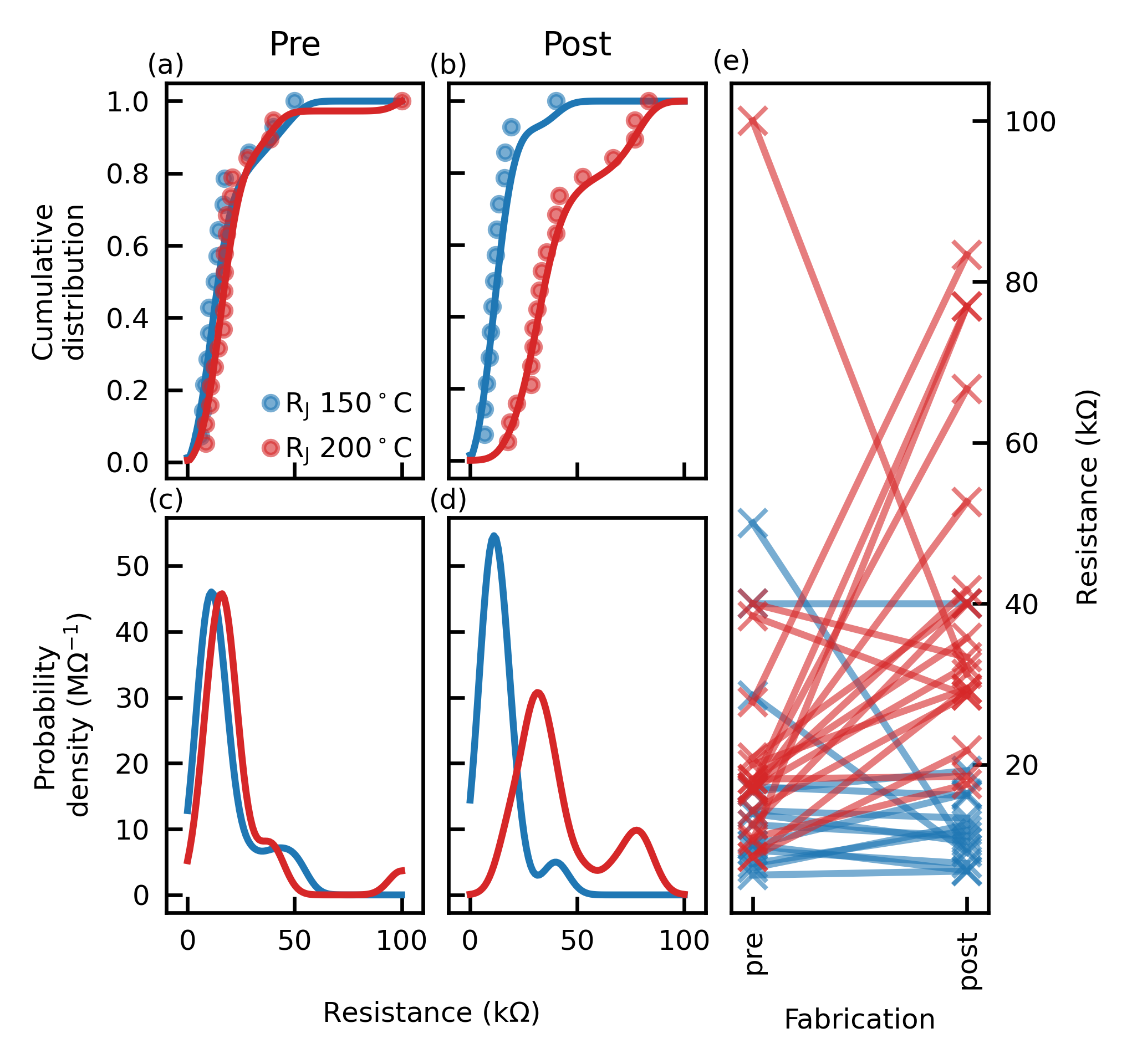}
\caption{
Temperature tests of two arrays of single SNS junctions that are exposed to either $150\Celsius$ (blue) and $200\Celsius$ (red) for $5~\minutes$ in PMGI. The tests simulate the temperature excursions of the GSL method and the traditional reflow method, respectively.
(a,b) CDFs of junction resistance (a) prior to and (b) following the temperature test.
(c,d) PDFs derived from the CDFs (c) prior to and (d) following the temperature test.
A clear shift toward higher resistances is observed for the $200\Celsius$ test.
(e) Comparison of each junction resistance before and after the test.
Note the relatively similar initial distributions of resistance and the different final distributions.
\label{fig:TemperatureDependence}
}
\end{figure}

\begin{figure}
\centering
\includegraphics{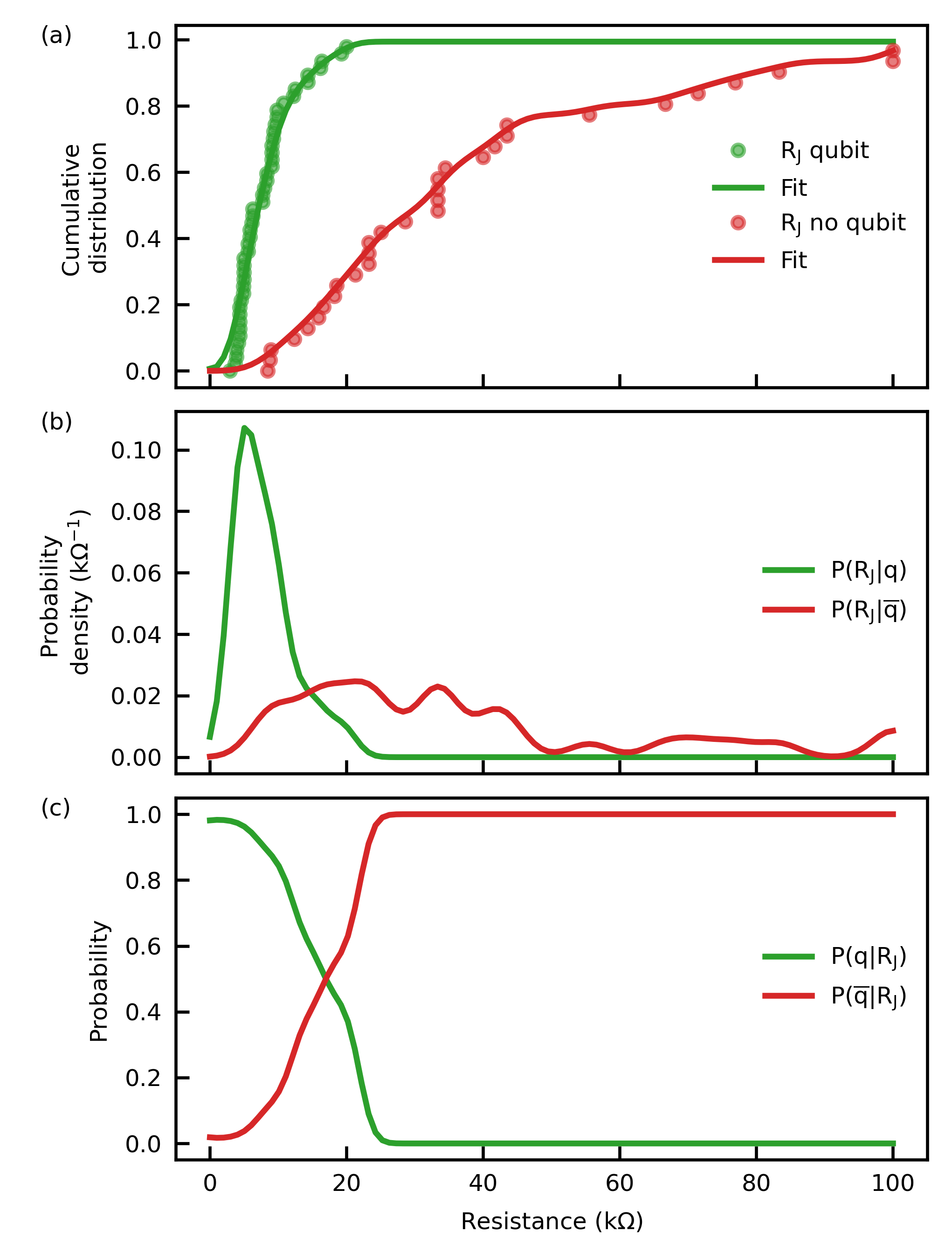}
\caption{
Study of the room-temperature resistance of the junction pairs in operable and non-operable SNS transmons.
(a) Cumulative distribution function of the resistance for operable (green) and non-operable (red) transmons.
Here, operable is conditioned on the observation of a power-dependent frequency shift in the dedicated readout resonator (see Fig.~\ref{fig:powershift} for an example).
(b) PDF derived from (a).
(c) Posterior probability [calculated from (b)]  of having an operable transmon as a function of its room-temperature JJ resistance.
\label{fig:QubitResistance}
}
\end{figure}

For a valid comparison, it is important that initial junction resistances for both chips be similar.
Two-point resistance measurements using a manual probe station confirm the overlap of cumulative distribution functions (CDFs) of initial resistance for both chips, as shown in Fig.~\ref{fig:TemperatureDependence}(a).
We perform a fit using kernel density estimation~\cite{Pedregosa11} to each of these CDFs and compute the derivative of the best fits to estimate the probability distribution functions (PDFs) of resistance. The results, shown in Fig.~\ref{fig:QubitResistance}(c), reveal a pre-test concentration around $20~\kOhm$ for both chips. The different temperature excursions make the resistance distributions become qualitatively different, as shown by the CDFs in Fig.~\ref{fig:TemperatureDependence}(b) and the PDFs in Fig.~\ref{fig:TemperatureDependence}(d) (similarly obtained).
For junctions exposed to $150\Celsius$ $(200\Celsius)$, the distribution of resistances shifts downward (upward).  The trajectory of individual junctions can be followed in Fig.~\ref{fig:TemperatureDependence}(e). For $150\Celsius$, the majority of resistances stay close to their initial values. For $200\Celsius$, however, the majority increase. Some junction resistances do decrease in both cases, particularly ones starting at the high end. While we do not understand the reason for this decrease, we speculate that it may arise from the different cleaning procedures used after the initial JJ contacting (see Supplementary Material) and after the simulated AB step.

Finally, we connect the of a transmon as a qubit at cryogenic temperature to the room-temperature resistance of its SNS junction pair.
We deem a transmon to be operable if we can simply observe of a power-dependent shift of the frequency of its readout resonator (see Fig.~\ref{fig:powershift} for an example). In total 78 qubits were measured from 8 different devices. These devices fall into three categories: 3 devices without ABs, in which 18 of 25 transmons were operable; 1 device with ABs fabricated by reflow, in which 1 of 9 transmons were operable;  and 4 devices with ABs fabricated by GSL, in which 28 of 44 transmons were operable. Figure~\ref{fig:QubitResistance}(a) shows numerical CDFs of the junction pair resistance for transmons that exhibit resonator power shifts (green) and for transmons that do not (red). These data clearly show that the resistance corresponding to an operable transmon is generally lower than that of a non-operable one. Fits to these numerical CDFs are done using kernel density estimation~\cite{Pedregosa11}.
The derivative of each best fit gives a probability density function (PDF) [Fig.~\ref{fig:QubitResistance}(b)].
Using a Bayesian update, we extract the posterior probability of a transmon being operable given its room-temperature resistance. The probability [Fig.~\ref{fig:QubitResistance}(c)] starts off close to unity and decreases to 0.5 by $\sim18~\kOhm$. The probability reduces to near zero by $\sim25~\kOhm$.
We conclude that for a good SNS Josephson junction it is vital that the room-temperature resistance be as low as possible, cementing the benefits of GSL-based AB fabrication.

In summary, we have employed grayscale lithography to reduce the peak temperature for airbridge processing compared to the traditional reflow method. We have shown that lowering peak processing temperature from $200\Celsius$ (needed for PMGI reflow) to $150\Celsius$ (limited by PMGI adhesion) increases the yield of operable SNS transmons based on InAs-nanowire Josephson junctions. We have done this in two steps. First we showed that GSL-based fabrication produces lower room-temperature JJ resistances. Secondly, we showed that lower JJ resistance increases the probability of having an operable SNS transmon at cryogenic temperature. For future work, it remains important to correlate the AB fabrication process with SNS transmon coherence time. It is also worthwhile to explore other e-beam resists that bake at lower temperatures without suffering adhesion problems as well as optical GSL using a direct laser writer, which could possibly lower baking even to room temperature.

\begin{acknowledgments}
We thank S.~A.~Khan and P.~Krogstrup for supplying the InAs nanowires, C.~Zachariadis for fabrication assistance, J.~Kroll and A.~Bruno for discussions, and C.~Andersen for comments on the manuscript.
This research is funded by the European Research Council (ERC) Synergy grant QC-lab and by the Allowance for Top Consortia for Knowledge and Innovation (TKIs) of the Dutch Ministry of Economic Affairs.

Correspondence and requests for materials should be addressed to L.D.C. (l.dicarlo@tudelft.nl)
The data shown in all figures of the main text and Supplementary Information are available at
\verb"http://github.com/DiCarloLab-Delft/"\\
\verb"Grayscale_Lithography".
\end{acknowledgments}

%\bibliographystyle{apsrev4-2}
%\putbib[../../Paper_resources/References/References_cQED]

%
\end{bibunit}
\clearpage

\renewcommand{\theequation}{S\arabic{equation}}
\renewcommand{\thefigure}{S\arabic{figure}}
\renewcommand{\thetable}{S\arabic{table}}
\renewcommand{\bibnumfmt}[1]{[S#1]}
\renewcommand{\citenumfont}[1]{S#1}
\setcounter{figure}{0}
\setcounter{equation}{0}
\setcounter{table}{0}

\begin{bibunit}[apsrev4-2]

\onecolumngrid
\section*{Supplementary material for `'Lower-temperature fabrication of airbridges by grayscale lithography to increase yield of nanowire transmons in circuit QED quantum processors''}
\FloatBarrier
This supplementary material describes the SNS junction fabrication, compares the processes for airbridge fabrication using the GSL method and the traditional reflow method (Table~\ref{tab:comparison} and Fig.~\ref{fig:powershift}), and shows a typical example of a power-dependent resonator frequency shift (Fig.~\ref{fig:powershift}).

\subsection*{SNS junction fabrication}
\label{sec:SNSJunction}
The SNS transmon fabrication recipe is adopted from~\cite{Luthi18}. First, the nanowire is transferred from a growth chip to the device using a nanomanipulator.
A $180~\nm$ thick layer of PMMA 950k is applied and baked for $5~\minutes$ on a hotplate at $150\Celsius$.
Using e-beam lithography, a $80~\nm$ rectangular window defined at the desired location of junction, where the Al is to be removed.
The PMMA is developed using a MIBK/IPA mixture with 1:3 volume ratio for $60~\seconds$, followed by a $10~\seconds$ dunk in an ethanol/IPA mixture with 1:3 volume ratio, and finally a $10~\seconds$ rinse in IPA.
The Al is etched using Transene D at $48.2\Celsius$ for $12~\seconds$, followed immediately by two dunks in water (first $5~\seconds$ and then $30~\seconds$).
The junction defining process is finished by removing the PMMA in acetone for $5~\minutes$ at $55\Celsius$ and cleaning with IPA for $10~\seconds$ at $55\Celsius$ followed by blow-drying.

To contact the nanowire junctions to the transmon capacitor pads, a $280~\nm$ layer of PMMA is spun and baked for $5~\minutes$ at $150\Celsius$. The e-beam writing and development is the same as for the etch windows. After development, the chip is loaded into a sputtering machine, where a $120~\nm$ thick layer of NbTiN is deposited. An in-situ argon mill is first done for $90~\seconds$ at $50~\Watt$ and $3~\mTorr$ to improve the contacting to the nanowire. (The duration of this critical process was pre-optimized for the lowest junction resistance.) Immediately afterwards, a thin NbTi sticking layer is deposited followed by the DC-sputtering of NbTiN at $2.5~\mTorr$ and $250~\Watt$.

\subsection*{Airbridge fabrication using the reflow method}
\begin{figure}
\centering
\includegraphics{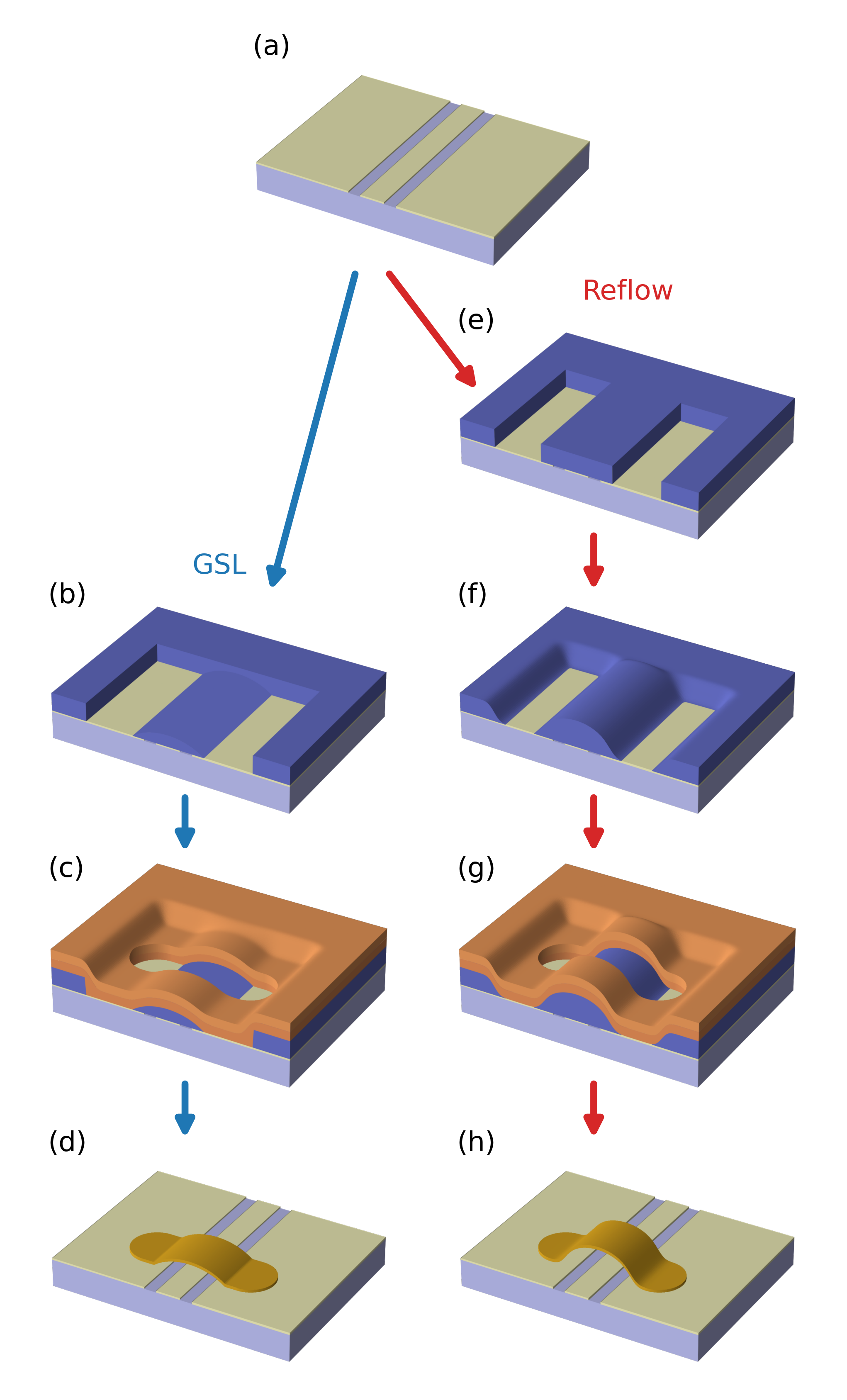}
\caption{
Comparison of airbridge fabrication steps using the GSL method (left, blue arrows) and the reflow method (right, red arrows). (a) Both methods start with the pre-fabrication of the base layer.  (b,e) A layer of PMGI is spun and developed for both methods. The GSL method directly produces the round profile. (f) The reflow method requires reflow at $200\Celsius$ to produce the round profile. (c,g) A PMMA bilayer is used to define the lateral airbridge dimensions. (d, h) NbTiN is sputtered and lifted off.
\label{fig:Reflow_grayscale_comp}
}
\end{figure}

The fabrication process for the reflow method starts following pre-patterning of the chip base layer. A $6.4~\um$ thick layer of PMGI SF15 is spun in 2 layer steps. Both layers are baked for $5~\minutes$ on a hotplate at $180\Celsius$. Then, using e-beam lithography, a rectangular profile with clearances is made at the desired position of airbridges. An AZ400k/water mixture in a 1:4 volume ratio is then use to develop the PMGI. The chip is dunked into the developer for $50~\seconds$, followed by a thorough water rinse for $30~\seconds$, and finished by blow-drying. The chip is then placed on a hotplate at $200\Celsius$ for $5~\minutes$ to reflow the resist and thus produce round profile. Due to surface tension, the resulting height of the PMGI at the airbridge location is higher than the original resist height. The resulting layer is shown in Fig.~\ref{fig:Reflow_grayscale_comp}(f). Next, a $400~\nm$ thick layer of PMMA 495K is spun and baked on a hotplate at $150\Celsius$ for $5~\minutes$, immediately followed by a $1.5~\um$ thick layer of PMMA 950k spun and baked in the same way. After e-beam lithography and development, the resist looks as in Fig.~\ref{fig:Fab_process}(g).

\begin{table}[ht]
\centering
% To place a caption above a table
\begin{tabular}[t]{|l|c|c|}
\hline
Requirement & GSL & Reflow\\
\hline
\hline
Resist contrast & low & any\\
Resist type & positive & positive or negative\\
Need compatibility with solvent of top top resist stack & yes & yes\\
Can developer top resist stack develop the bottom resist?& no & no\\
\hline
\end{tabular}
% Or to place a caption below a table
\caption{Comparison of the requirements for the resist used for the GSL method and the traditional reflow method.}
\label{tab:comparison}
\end{table}%

\subsection*{Resonator power-induced frequency shift}
\begin{figure}
  \centering
    \includegraphics{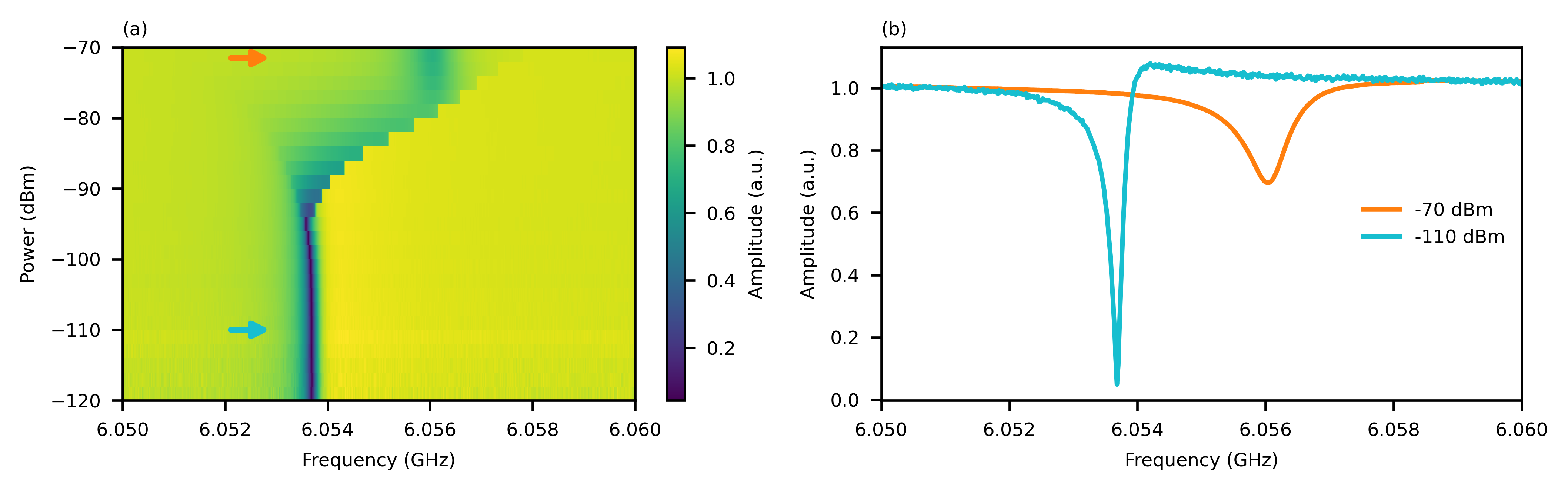}
    \caption{Example shift of the resonance frequency of a dedicated readout resonator with increasing incident power on the feedline, indicating that the coupled SNS transmon is operable. (a) Image plot of normalized feedline transmission as a function of probe frequency and incident power. The resonance shifts from $6.0538~\GHz$ at $-110~\dBm$ to $6.0560~\GHz$ at $-70~\dBm$. This positive shift indicates that the qubit transition frequency is above the resonator frequency. (b) Linecuts of feedline transmission versus frequency at $-110~\dBm$ (cyan) and $-70~\dBm$ (orange).}
    \label{fig:powershift}
\end{figure}

We judge whether or not a transmon is operable by determining whether its dedicated readout resonator exhibits a frequency shift when measured with increasing incident power. A typical measurement of a readout resonator as a function of incident power on the feedline is shown in Fig.~\ref{fig:powershift}. In this case, there is a upward $2.2~\MHz$ shift of the resonance frequency with increasing power. A positive (negative) frequency shift indicates that the transmon qubit transition frequency lies above (below) that of the resonator. For SNS transmons based on InAs nanowires, the qubit transition frequency cannot be accurately targeted, and can fall above and below the resonator.

%\bibliographystyle{apsrev4-2}
%\putbib[../../Paper_resources/References/References_cQED]

%
\end{bibunit}

\end{document}